\documentstyle[11pt]{article}

\newcommand{\be}{\begin{equation}}
\newcommand{\ee}{\end{equation}}
\newcommand{\bea}{\begin{eqnarray}}
\newcommand{\eea}{\end{eqnarray}}
\newcommand{\bref}[1]{(\ref{#1})}
\newcommand{\eps}{\epsilon}
\newcommand{\veps}{\varepsilon}

\newcommand{\der}[2]{\frac{\partial #1}{\partial #2}}
\newcommand{\gh}[1]{{\cal #1}}
\newcommand{\agh}[1]{\bar{\cal #1}}

\makeatletter

\def\mathoperat{\@ifnextchar [{\@mathoperat}{\@mathoperat[rm]}}
   \def\@mathoperat[#1]#2#3{\def#2{\mathop{\@nameuse{#1} #3{}}\nolimits}}

\def\restric#1#2{{\left. #1 \right|_{#2}}}
\def\dif{{\rm d}}
\def\deriv{\@ifnextchar[{\@deriv}{\@deriv[]}}
   \def\@deriv[#1]#2#3{\mathchoice%
{{\dif^{#1}#2\over\dif{#3}^{#1}}}{{\dif^{#1}#2/\dif{#3}^{#1}}}%
{{\dif^{#1}#2\over\dif{#3}^{#1}}}{{\dif^{#1}#2/\dif{#3}^{#1}}}}

\def\secteqno{\@addtoreset{equation}{section}%
\def\theequation{\thesection.\arabic{equation}}}

\def\endsecteqno{\def\theequation{\@ifundefined{chapter}%
{\arabic{equation}}{\thechapter.\arabic{equation}}}}

\newcounter{subequation}
\def\thesubequation{\alph{subequation}}
\def\sneqnarray{\stepcounter{equation}\let\@currentlabel=\theequation
\setcounter{subequation}{1}
\def\@eqnnum{{\rm (\theequation.\thesubequation)}}
\global\@eqcnt\z@\tabskip\@centering\let\\=\@eqncr\let\@@eqncr=\@@sneqncr
$$\halign to \displaywidth\bgroup\@eqnsel\hskip\@centering
 $\displaystyle\tabskip\z@{##}$&\global\@eqcnt\@ne
 \hskip 2\arraycolsep \hfil${##}$\hfil
 &\global\@eqcnt\tw@ \hskip 2\arraycolsep $\displaystyle\tabskip\z@{##}$\hfil
  \tabskip\@centering&\llap{##}\tabskip\z@\cr}
\def\endsneqnarray{\@@sneqncr\egroup $$\global\@ignoretrue}
\def\@@sneqncr{\let\@tempa\relax
   \ifcase\@eqcnt \def\@tempa{& & &}\or \def\@tempa{& &}
   \else \def\@tempa{&}\fi
     \@tempa \if@eqnsw\@eqnnum\stepcounter{subequation}\fi
     \global\@eqnswtrue\global\@eqcnt\z@\cr}

\def\nobiblabels{\def\@lbibitem[##1]##2{\@bibitem{##2}}}

\makeatother

\title{\bf Faddeev-Popov method for anomalous quasigroups}

\author{\sc{Jordi Par\'{\i}s}\thanks{Bitnet address: PARIS@EBUBECM1}\\
        \small{\it{Departament d'Estructura i Constituents
               de la Mat\`eria}}\\
        \small{\it{Universitat de Barcelona}}\\
        \small{\it{Diagonal 647}}\\
        \small{\it{E-08028 BARCELONA}}\\
        \small{\it{Catalonia, Spain}}}

\date{}

\begin{document}
\maketitle
\thispagestyle{empty}

\begin{abstract}

\end{abstract}

\vfill %\hfill
\vbox{\raggedleft
 October 1992\par
 Universitat de Barcelona preprint UB-ECM-PF 92/24\par\null}

\clearpage

\newpage

\secteqno

\section{Introduction}

\hspace{\parindent}%
The quantization of anomalous gauge theories has received much attention
during the last few years. Following the work of Faddeev and Shatashvili
\cite{FS86}, some proposals appeared \cite{BSV86} \cite{HT87}
for cancelling the anomalies through Wess-Zumino terms \cite{WZ71}. The
main
observation of these works was that, in the path integral quantization
of chiral gauge theories, the integration over the elements of the gauge
group must be taken into account. In this way additional fields as
well as the Wess-Zumino action emerge naturally in the process of
quantization. A similar observation was made earlier by Polyakov
\cite{Pol81} in the path integral quantization of the bosonic string in
the presence of the conformal anomaly.

On the other hand, a proposal to deal with anomalous gauge theories with
closed, irreducible gauge algebra in the framework of the
field-antifield formalism was described recently in \cite{GP92}. There,
by extending the original configuration space with extra degrees of
freedom, a general expression for the Wess-Zumino action in terms of the
anomalies of the theory was obtained and the form of the gauge
transformations for the extra fields was given.

The aim of this letter is to describe a generalization of the
Faddeev-Popov procedure \cite{FP69} in order to treat on the same
footing the
path integral quantization of either anomalous and non-anomalous gauge
theories with closed, irreducible gauge algebra. Our construction
closely follows the results of ref. \cite{B80}, where the quantization
{\it a la} Faddeev-Popov of non-anomalous quasigroups was
considered. As a byproduct,
we obtain expressions for the Wess-Zumino action and for the gauge
transformations of the additional fields which are in complete agreement
with the ones previously given in \cite{GP92}.

Prior to that, since along our development some technical tools about
the so-called quasigroup structure
are needed, a summary of the results given in \cite{B80} is presented.

\section{Brief overview of Quasigroups}

\hspace{\parindent}%
The quasigroup structure is a generalization of the Lie group structure,
introduced by Batalin \cite{B80}. The main difference between this
structure and a Lie group relies on the fact that the quasigroup
composition law depends not only on the parameters of the
transformations, but also on the variables on which these
transformations act.

To introduce the concept of quasigroup, it is useful to have in
mind a manifold $\gh M$ with coordinates $\phi^i$, $i=1,\ldots,n$. Under
these conditions, consider
a continuous transformation acting on the coordinates of $\gh M$ given
by
$$
   \bar\phi^i=F^i(\phi,\theta),
$$
with $\theta^\alpha$, $\alpha=1,\ldots,r$, a set of real
parameters\footnote{Along this paper, for simplicity, we will restrict
          ourselves to the bosonic case, i.e.,
          $\eps(\phi^i)=\eps(\theta^\alpha)=0$.}.

Assume now that the transformations $F^i(\phi,\theta)$ satisfy the
following properties:
\begin{enumerate}
\item
For $\theta^\alpha=0$, we have the identity transformation
$$
   \phi^i=F^i(\phi,0).
$$
\item
The composition law between two finite transformations reads
\be
   F^i(F(\phi,\theta),\theta')=F^i(\phi,\varphi(\theta,\theta';\phi)),
\label{comp law}
\ee
where $\varphi^\alpha(\theta,\theta';\phi)$ is the
composition law of the quasigroup.
\item
 Left and right units coincide
\be
   \varphi^\alpha(\theta,0;\phi)=\varphi^\alpha(0,\theta;\phi)=
   \theta^\alpha.
\label{units}
\ee
\item
A modified associativity law holds
\be
   \varphi^\alpha(\varphi(\theta,\theta';\phi),\theta'';\phi)=
\varphi^\alpha(\theta,\varphi(\theta',\theta'';F(\phi,\theta));\phi).
\label{ass law}
\ee
\item
There exists an inverse transformation given by
$$
   \phi^i=F^i(\bar\phi,\tilde\theta(\theta,\bar\phi)),
$$
with the inverse $\tilde\theta^\alpha(\theta,\bar\phi)$ satisfying the
relations
$$
   \varphi^\alpha(\tilde\theta(\theta,\bar\phi),\theta;\bar\phi)=
   \varphi^\alpha(\theta,\tilde\theta(\theta,\bar\phi);\phi)=0.
$$
\end{enumerate}

The above conditions 1)-5) define the quasigroup structure.
Some of the relations describing the quasigroup at the infinitesimal
level, which are useful in the present study, can be obtained from these
conditions in the following way.

The generators of infinitesimal transformations for the
variables $\phi^i$ are
$$
  R^i_\alpha(\phi)\equiv\restric{\frac{\partial F^i(\phi,\theta)}
  {\partial \theta^\alpha}}{\theta=0}.
$$
Antisymmetrizing the second derivatives of the modified composition law
(\ref{comp law}) with respect to $\theta^\alpha$, $\theta^{'\beta}$ in
$\theta=\theta'=0$, one obtains the algebra
\be
    R^i_{\alpha,j}R^j_{\beta}-R^i_{\beta,j}R^j_{\alpha}=
    R^i_{\gamma}T^{\gamma}_{\alpha\beta},
\label{gauge algebra a}
\ee
with the structure functions $T^\gamma_{\alpha\beta}(\phi)$ defined as
$$
  T^\gamma_{\alpha\beta}(\phi)=\restric{
  \left[\frac{\partial^2\varphi^\gamma(\theta,\theta';\phi)}
  {\partial\theta^\beta\partial\theta^{'\alpha}}-(\alpha,\beta)\right]}
  {\theta=\theta'=0}.
$$

{}From the associativity law (\ref{ass law}), one gets
the generalized Jacobi identity for the
structure functions $T^\gamma_{\alpha\beta}$
\be
    T^{\mu}_{\alpha\gamma}T^{\gamma}_{\beta\delta}-
    T^{\mu}_{\alpha\beta,i}R^i_{\delta}+
    (\mbox{cyclic perm. of ($\alpha$, $\beta$, $\delta$}))=0.
\label{jacobi a}
\ee

Let us introduce now the matrices $\mu$, $\tilde\mu$ defined by
\be
   \mu^\alpha_\beta(\theta,\phi)=\restric{
   \frac{\partial\varphi^\alpha(\theta,\theta';\phi)}
   {\partial\theta^{'\beta}}}{\theta'=0},\quad\quad
   \tilde \mu^\alpha_\beta(\theta,\phi)=\restric{
   \frac{\partial\varphi^\alpha(\theta',\theta;\phi)}
   {\partial\theta^{'\beta}}}{\theta'=0},
\label{mu}
\ee
while we denote their inverses as $\lambda$ and $\tilde\lambda$
$$
  \lambda^\alpha_\gamma\mu^\gamma_\beta=\delta^\alpha_\beta,
   \quad\quad\quad
  \tilde\lambda^\alpha_\gamma\tilde\mu^\gamma_\beta=\delta^\alpha_\beta.
$$
These matrices will always exist, at least locally,
because the property
$$
  \restric{\mu^\alpha_\beta}{\theta=0}=
  \restric{\tilde\mu^\alpha_\beta}{\theta=0}=\delta^\alpha_\beta,
$$
holds by virtue of eqs.(\ref{units}) and \bref{mu}.

With the aid of the associativity law (\ref{ass law})
one obtains an analog to the Lie equation for the transformations of
the fields $F^i$
\be
   \frac{\partial F^i}{\partial\theta^\alpha}=
   R^i_\beta(F)\lambda^\beta_\alpha(\theta,\phi),
\label{f derivada}
\ee
and the useful transformation rule for the generators
\be
   \frac{\partial F^i}{\partial\phi^j}R^j_\beta(\phi)=
R^i_\gamma(F)\lambda^\gamma_\alpha\tilde\mu^\alpha_\beta(\theta,\phi).
\label{trans gen}
\ee

On the other hand, differentiating the same equation
\bref{ass law} with respect to the parameters of the quasigroup,
an analog to the Lie equation for the composition functions
$\varphi^\alpha$ is found
\be
  \frac{\partial\varphi^\alpha(\theta,\theta';\phi)}
  {\partial\theta^{'\gamma}}
  =\mu^\alpha_\beta(\varphi(\theta,\theta';\phi),\phi)
    \lambda^\beta_\gamma(\theta',F(\phi,\theta)),
\label{ultima}
\ee
as well as the following commutation relation for the elements of the
matrix $\lambda$
\be
    \der{\lambda^\alpha_\gamma}{\theta^\beta}-
    \der{\lambda^\alpha_\beta}{\theta^\gamma}-
    T^\alpha_{\mu\nu}(\bar\phi)
    \lambda^\mu_\beta\lambda^\nu_\gamma=0,
\label{algebra lambda}
\ee
which is the analog to the Maurer-Cartan equation for a Lie group.

Another useful relation involving the matrices $\lambda$,
$\tilde\lambda$, $\mu$, $\tilde\mu$ is
\be
   \lambda^\delta_\gamma(D_\beta\mu^\alpha_\delta)-
\tilde\lambda^\delta_\beta\frac{\partial\tilde\mu^\alpha_\delta}
   {\partial\theta^\gamma}=0,
\label{mu lambda t}
\ee
with the operator $D_\beta$ defined as
$$
   D_\beta\equiv\left(\frac{\partial}{\partial\theta^\beta}-
   R^i_\alpha(\phi)\tilde\lambda^\alpha_\beta(\theta,\phi)
   \frac{\partial}{\partial\phi^i}\right).
$$

In our construction some functional integrals over the elements of
the gauge quasigroup should be considered. For this reason
it is convenient to introduce the concept of right and left
invariant measures on the quasigroup. The right invariant measure,
defined by
\be
    \gh D G_R(\theta,\phi)\equiv\left[\gh D \theta
    \det\tilde\lambda^\alpha_\beta(\theta, F(\phi,\tilde\theta))
    \right],
\label{right m}
\ee
is invariant under the simultaneous change of classical fields and
parameters
\bea
     &\phi^i\rightarrow\bar\phi^i=F^i(\phi,\veps),&
\nonumber\\
    &\theta^\alpha\rightarrow\bar\theta^\alpha=
    \varphi^\alpha(\theta,\veps;F(\phi,\tilde\theta)),&
\nonumber
\eea
where $\tilde\theta^\alpha=\tilde\theta^\alpha(\theta,\bar\phi)$ is the
inverse of the parameter $\theta^\alpha$.

On the other hand, the left invariant measure
\be
    \gh D G_L(\theta,\phi)\equiv
    \left[\gh D \theta \det\lambda^\alpha_\beta(\theta,\phi)\right],
\label{left m}
\ee
can be shown to be invariant under the simultaneous change of classical
fields and parameters
\bea
     &\phi^i\rightarrow\bar\phi^i=F^i(\phi,\veps),&
\nonumber\\
    &\theta^\alpha\rightarrow\bar\theta^\alpha=
    \varphi^\alpha(\tilde\veps(\veps,\bar\phi),\theta;
    \bar\phi).&
\label{trans theta}
\eea

In particular, the invariance property of the left invariant measure
turns out to be very important in our construction. Indeed, as we will
see, for the case of anomalous gauge theories, this invariance suggests
considering the parameters of the quasigroup as new dynamical variables
transforming under the action of the gauge quasigroup as in
\bref{trans theta}.

For a more exhaustive study of the quasigroup structure,
we refer the reader to the original reference \cite{B80}.

\section{Generalized Faddeev-Popov method for quasigroups}

\hspace{\parindent}%
Let us now describe a generalization of the Faddeev-Popov procedure
\cite{FP69} to treat on the same footing the path integral quantization
of either anomalous and non-anomalous gauge theories.

We restrict ourselves to irreducible theories with closed gauge
algebras. The classical action for these systems,
$S_0(\phi^i)$, $i=1,\ldots,n$, is invariant under the
(infinitesimal) gauge transformations
\be
   \delta\phi^i=R^i_\alpha(\phi)\veps^\alpha,\quad\quad\quad
   \alpha=1,\ldots,r.
\label{trans fi}
\ee
Besides, the generators $R^i_\alpha$ are assumed to be independent and
the relations (\ref{gauge algebra a}) and (\ref{jacobi a}) verified
at any space-time point. Therefore, the gauge structure is in
general that of the quasigroup defined in the preceding section at
every space-time point.

In order to generalize the Faddeev-Popov procedure for a generic
quasigroup let us consider, first of all, admissible gauge fixing
conditions $\chi_\alpha(\phi)=0$. After that, a representation of the
unity for quasigroups is introduced \cite{B80}
\be
   1_L=\int{\cal D}\theta\,\delta[\chi_\alpha(F(\phi,\theta))]\,
   \Delta_\chi[F(\phi,\theta)].
\label{left in measure}
\ee
The determinant $\Delta_\chi[F(\phi,\theta)]$ in
\bref{left in measure}, given by
$$
   \Delta_\chi[F(\phi,\theta)]=
   \det\left(\frac{\partial\chi_\alpha(F(\phi,\theta))}
   {\partial\theta^\beta}\right),
$$
can be written, using relation \bref{f derivada}, as
$$
   \Delta_\chi[F(\phi,\theta)]=
   \left[\det D_{\alpha\beta}(F(\phi,\theta))\right]
    \left[\det\lambda^\alpha_\beta(\theta,\phi)\right],
$$
where the matrix $D_{\alpha\beta}$ is defined through the relation
$$
   D_{\alpha\beta}(\phi)=
   \left(\frac{\partial\chi_\alpha}{\partial\phi^i}R^i_\beta
    \right)(\phi).
$$

Hence, the above expressions yield the following representation of the
unity $1_L$ (\ref{left in measure})
$$
   1_L=\int\left[{\cal D}\theta
   \det\lambda^\alpha_\beta(\theta,\phi)\right]\,
   \delta[\chi_\alpha(F(\phi,\theta))]\,
   \left[\det D_{\alpha\beta}(F(\phi,\theta))\right],
$$
where $[\det D_{\alpha\beta}(\phi)]$ is the usual Faddeev-Popov
determinant and
$$
    \left[{\cal D}\theta\det\lambda^\alpha_\beta(\theta,\phi)\right]
    \equiv\gh D G_L(\theta,\phi),
$$
is the left invariant measure for the elements of the quasigroup
\bref{left m}.

An equivalent representation of the unity, $1_R$, can be obtained as
well from \bref{left in measure} through the change of parameters
$\theta^\alpha\rightarrow\tilde\theta^\alpha(\theta,\bar\phi)$,
\be
   1_R=\int\left[{\cal D}\theta
   \det\tilde\lambda^\alpha_\beta(\theta,F(\phi,\tilde\theta))\right]
   \delta[\chi_\alpha(F(\phi,\tilde\theta))]\,
   \left[\det D_{\alpha\beta}(F(\phi,\tilde\theta))\right],
\label{right in measure}
\ee
where $\tilde\theta^\alpha=\tilde\theta^\alpha(\theta,\bar\phi)$
is the inverse element associated with $\theta^\alpha$ and
$$
   \left[{\cal D}\theta
   \det\tilde\lambda^\alpha_\beta(\theta,F(\phi,\tilde\theta))\right]
   \equiv \gh D G_R(\theta,\phi),
$$
is now the right invariant measure for the elements of the
quasigroup \bref{right m}.

Consider now
the vacuum-to-vacuum transition amplitude or S-matrix of the theory
\be
   Z=\int{\cal D}\phi\,\exp\left\{\frac{i}{\hbar} S_0(\phi)\right\},
\label{smatrix}
\ee
where $\gh D\phi$ is the naive integration measure.
After inserting in it the unity $1_R$ \bref{right in measure} in the
usual way, it reads
\bea
   Z&=&\int{\cal D}\phi\,[{\cal D}\theta
   \det\tilde\lambda^\alpha_\beta(\theta,F(\phi,\tilde\theta))]\,
   \delta[\chi_\alpha(F(\phi,\tilde\theta))]
\nonumber\\
  &&\hspace{8mm}\left[\det D_{\alpha\beta}(F(\phi,\tilde\theta))\right]
  \exp\left\{\frac{i}{\hbar} S_0(\phi)\right\}.
\nonumber
\eea
The dependence of $D_{\alpha\beta}$ and $\chi_\alpha$ on $\theta^\alpha$
can be dropped out by performing the change of variables
$\phi^i\rightarrow F^i(\phi,\theta)$. This operation yields
\be
   Z=\int{\cal D}F(\phi,\theta)\,
  [{\cal D}\theta\,\det\tilde\lambda^\alpha_\beta(\theta,\phi)]\,
  \delta[\chi_\alpha(\phi)] \left[\det D_{\alpha\beta}(\phi)\right]
  \exp\left\{\frac{i}{\hbar} S_0(\phi)\right\}.
\label{z}
\ee

At this point, in the standard Faddeev-Popov procedure it is assumed
that the functional measure is gauge invariant. However, as is well
known after Fujikawa's works \cite{Fuj}, this is not true in general,
this non-invariance being the source of potential anomalies. Therefore,
a careful analysis of the jacobian of the above change of variables
\be
   {\cal D}F(\phi,\theta)= \det\left(\der{F^i}{\phi^j}\right)
   {\cal D}\phi=[\det S^i_{\,j}]\,{\cal D}\phi,
\label{jacobia}
\ee
should be considered.

An expression for this jacobian
can be obtained, as described in \cite{B80}, by differentiating its
logarithm with respect to the parameters of the quasigroup. Proceeding
in this way and taking into account eqs.\bref{f derivada} and
\bref{trans gen} we find
$$
  \der{}{\theta^\alpha}\left\{\ln[\det S^i_{\,j}]\right\}=
   R^i_{\beta,i}(F)\lambda^\beta_\alpha-
   \mu^\sigma_{\beta,i} R^i_\gamma(\phi)\tilde\lambda^\gamma_\sigma
   \lambda^\beta_\alpha.
$$
The last term of the above expression can be written, using
\bref{algebra lambda} and \bref{mu lambda t}, as
$$
   \mu^\sigma_{\beta,i} R^i_\gamma(\phi)\tilde\lambda^\gamma_\sigma
   \lambda^\beta_\alpha=- T^\gamma_{\gamma\beta}(F)\lambda^\beta_\alpha-
   \der{}{\theta^\alpha}
   \left\{\ln\left(\frac{\det\lambda}{\det\tilde\lambda}\right)\right\},
$$
from which we obtain
\be
  \der{}{\theta^\alpha}\left\{\ln[\det S^i_{\,j}]\right\}=
   A_\beta(F)\lambda^\beta_\alpha+
   \der{}{\theta^\alpha}
   \left\{\ln\left(\frac{\det\lambda}{\det\tilde\lambda}\right)\right\},
\label{ldet}
\ee
with the quantities $A_\alpha(\phi)$ given by
\be
  A_\alpha(\phi)=(R^i_{\alpha, i}+T^\beta_{\beta\alpha})(\phi).
\label{anomalia}
\ee
It should be noted that for a local gauge theory the quantities
$A_\alpha$ are proportional to $\delta(0)$. Therefore,
in order to make sense out of this construction, some scheme
to regularize them should be considered\footnote{From now on, $A_\alpha$
will stand for the regularized expression of \bref{anomalia}.}.

Introduce now an object $M_1(\phi,\theta)$ verifying the differential
equation
\be
  \frac{\partial M_1}{\partial\theta^\alpha}=
  -i A_\beta(F(\phi,\theta))\lambda^\beta_\alpha(\theta,\phi),
\label{eq m}
\ee
with the boundary condition $M_1(\phi,\theta=0)=0$ (mod $2\pi$). This
definition allows to solve immediately eq.\bref{ldet} and to write the
jacobian \bref{jacobia} as
$$
   [\det S^i_{\,j}](\phi,\theta)=\exp\{i M_1(\phi,\theta)\}
   \left(\frac{\det\lambda(\theta,\phi)}{\det\tilde\lambda(\theta,\phi)}
   \right),
$$
from which we obtain, substituting it into \bref{z}, the following
expression for the S-matrix
$$
   Z=\int{\cal D}\phi\,{\cal D}G_L(\theta,\phi)\,
   \delta[\chi_\alpha(\phi)]\left[\det D_{\alpha\beta}(\phi)\right]\,
   \exp\left\{\frac{i}{\hbar}[S_0(\phi)+
   \hbar M_1(\phi,\theta)]\right\}.
$$

Finally, the introduction of ghosts $\gh C^\alpha$, antighosts $\agh
C^\alpha$ and auxiliary fields $B^\alpha$ enables to exponentiate the
Faddeev-Popov
determinant and the gauge fixing conditions in the usual way, yielding
\be
   Z=\int
   \gh D\phi\,\gh D\gh C\,\gh D\agh C\,\gh D B\,\gh D G_L(\theta,\phi)
   \,\exp\left\{\frac{i}{\hbar}[S_\chi(\phi;\gh C,\agh C,B)+
   \hbar M_1(\phi,\theta)]\right\},
\label{z2}
\ee
where $S_\chi$ is the quantum gauge fixed action
$$
  S_\chi=S_0+\agh C^\alpha D_{\alpha\beta}\gh C^\beta+
         B^\alpha\chi_\alpha,
$$
and $\gh D G_L(\theta,\phi)$ is the left invariant measure for the
elements of the
gauge quasigroup (\ref{left m}).  Therefore, apart from the
explicit integration over the elements of the gauge quasigroup and
the presence of the $M_1$ term in \bref{z2}, the result we arrive for
the S-matrix is the same as the one
obtained in the standard Faddeev-Popov method.

Let us now consider in more detail the $M_1$ term. As is well known, the
integrability conditions for an equation of the type \bref{eq m} are
$$
   \frac{\partial^2 M_1}{\partial\theta^\alpha\partial\theta^\beta}-
   \frac{\partial^2 M_1}{\partial\theta^\beta\partial\theta^\alpha}=0.
$$
In the present case, after using eqs.\bref{f derivada} and
\bref{algebra lambda}, these conditions yield
$$
  \left(\der{A_\alpha}{\phi^i} R^i_\beta-
  \der{A_\beta}{\phi^i} R^i_\alpha-
  A_\gamma T^\gamma_{\alpha\beta}\right)(F)
  \lambda^\alpha_\mu\lambda^\beta_\nu=0,
$$
which, taking into account the invertibility of the matrix
$\lambda^\alpha_\beta$,
are equivalent to the Wess-Zumino consistency conditions \cite{WZ71}
\be
  \left(\der{A_\alpha}{\phi^i} R^i_\beta-
  \der{A_\beta}{\phi^i} R^i_\alpha-
  A_\gamma T^\gamma_{\alpha\beta}\right)(\phi)=0.
\label{wzcc}
\ee
Hence, from this result we conclude that the above construction makes
sense if the scheme introduced to regularize the
quantitites $A_\alpha$ \bref{anomalia} is ``consistent'', i.e.,
if the regularized expression of $A_\alpha$ verifies the Wess-Zumino
consistency conditions \bref{wzcc}.

Under such assumptions, the solution of
equation \bref{eq m} with the appropriate boundary condition is given by
the integral
\be
   M_1(\phi,\theta)=-i \int_0^\theta A_\beta
   (F(\phi,\theta'))\lambda^\beta_\alpha(\theta',\phi)
   \dif\theta'^\alpha,
\label{wesszumino}
\ee
which, in view of the above discussion, does not depend
upon the form of the integration path. Taking this fact into account and
using the analog to the Lie equation for the composition functions
\bref{ultima}, as well as \bref{comp law} and \bref{units}, it is
possible to verify that expression \bref{wesszumino} for the $M_1$ term
fulfills the so-called 1-cocycle condition
$$
  M_1(\phi,\theta)+  M_1(F(\phi,\theta),\theta')-
  M_1(\phi,\varphi(\theta,\theta';\phi))=0\quad(\mbox{mod}\;2\pi).
$$

Therefore, we conclude that the quantitites $A_\alpha$
\bref{anomalia} and the $M_1$ term \bref{wesszumino} are the
candidates to be the anomalies and the Wess-Zumino term
for the case of an anomalous gauge
theory. Note also that their expressions as well as the
expression \bref{z2} for the S-matrix coincide with
the ones obtained in \cite{GP92} in the framework of the field-antifield
formalism.

\section{Anomalous gauge theories}

\hspace{\parindent}%
The analysis performed so far has been given in full generality.
Let us now analyze in which conditions the theory can be
considered to be anomalous or not. To this end consider expression
\bref{z2} for the S-matrix and perform the integration over the elements
of the quasigroup. Once this is done, we have in general
$$
   \exp\{i \tilde M_1(\phi)\}=\int\gh D G_L(\theta,\phi)\,
   \exp\{i M_1(\phi,\theta)\}.
$$
After that, with the expression for the $\tilde M_1$ term at hand, we
face two possibilities:

\begin{itemize}

\item a) $\tilde M_1(\phi)$ is a local functional. In this case
$\exp\{i \tilde M_1(\phi)\}$ can be considered as part of the measure
of the fields of the theory and absorbed in it through a suitable
redefinition of them. This is the situation which corresponds to a
non-anomalous gauge theory.

Within this situation, it is of interest to consider
the particular case when $M_1$ is zero and the gauge group is a Lie
group. Under such conditions
the volume of the gauge group, $\int\gh D G_L(\theta)=V_{\rm gauge}$,
can be factorized from the functional integration, yielding as a
physically relevant quantity
$$
   Z_\chi=
   \int\gh D\phi\,\gh D\gh C\,\gh D\agh C\,\gh D B\,
   \exp\left\{\frac{i}{\hbar}S_\chi(\phi;\gh C,\agh C,B)\right\},
$$
obtaining in this way the standard Faddeev-Popov result for
non-anomalous gauge groups.

\item b) $\tilde M_1(\phi)$ is a non-local functional. This is the way
in which anomalous gauge theories manifest themselves in this
formulation. In this case in order to use standard techniques of
quantum field theory is better to take expression \bref{z2} as a
starting point. $\theta^\alpha$ are then interpreted as the additional
fields which appear
in anomalous gauge theories due to the loss of the gauge symmetry at the
quantum level and the $M_1$ term as the corresponding Wess-Zumino
action.

\end{itemize}

In the case of an anomalous gauge
theory, the invariance property of the left measure \bref{left m}
for the elements of the gauge quasigroup suggests taking
\bref{trans theta} as the gauge transformation for the extra fields
$\theta^\alpha$. It is also of interest to consider the infinitesimal
form of this transformation which reads
\be
   \delta\theta^\alpha=-\tilde\mu^\alpha_\beta(\theta,\phi)\veps^\beta.
\label{inf t}
\ee

To complete the construction, it is instructive
to verify that with the above choice
for the gauge transformations of the additional fields $\theta^\alpha$
the (infinitesimal) gauge variation
of the Wess-Zumino action \bref{wesszumino} reproduces the anomalies
$A_\alpha$. Indeed, using the infinitesimal transformations
\bref{trans fi} and \bref{inf t}, it is
\bea
  \delta M_1(\phi,\theta)&=&
  -i \int_0^\theta \delta[A_\beta
   (F(\phi,\theta'))\lambda^\beta_\alpha(\theta',\phi)]\dif\theta'^\alpha
\nonumber\\
   &&+i A_\sigma(F(\phi,\theta))\lambda^\sigma_\beta
     \tilde\mu^\beta_\gamma(\theta,\phi)\veps^\gamma,
\label{variacio m}
\eea
where the second term comes from the variation of the upper integration
limit. On the other hand, the gauge variation of the integrand yields a
total derivative
\be
   \delta[A_\beta(F(\phi,\theta'))\lambda^\beta_\alpha(\theta',\phi)]=
   \der{}{\theta^{'\alpha}}\left[
   A_\sigma(F(\phi,\theta'))\lambda^\sigma_\beta
     \tilde\mu^\beta_\gamma(\theta',\phi)\veps^\gamma\right],
\label{bt}
\ee
where in evaluating this expression use has been made of the
consistency conditions for $A_\alpha$ \bref{wzcc} and the commutation
relation for $\lambda^\alpha_\beta$ \bref{algebra lambda}. Finally, the
value of
the boundary term \bref{bt} in the upper limit of the integral exactly
cancels the second term in \bref{variacio m}, obtaining in the end
the expected result
$$
  \delta M_1(\phi,\theta)=  i A_\alpha(\phi)\veps^\alpha.
$$

In summary, we have described a generalization of the Faddeev-Popov
method to treat either anomalous and non-anomalous gauge theories in an
unified way. As a byproduct, explicit expressions for the Wess-Zumino
action in terms of the anomalies of the theory $A_\alpha$ as well as for
the gauge transformations of the extra fields have been obtained, in
complete agreement with the results presented in \cite{GP92}.

Finally, we would like to comment about the relation of our results
with the ones presented in \cite{OP}, where a derivation of the
Faddeev-Popov formula for a generic non-linear sigma model was
performed.
Although both results are equivalent in the non-anomalous case, the
procedure followed in this paper is different from ours. Indeed, while
we start from the naive S-matrix \bref{smatrix}, in \cite{OP} an
S-matrix with a reparametrization invariant measure, including an extra
factor, is considered. On the other hand, a major difference comes from
the fact that the representation of the unity considered there is
constructed in terms of a path integral along the gauge orbits, without
any reference to the gauge group. Clearly
a further study is needed to elucidate the relationship between both
procedures.

\section{Example: The chiral Schwinger model}

\hspace{\parindent}%
To conclude, let us apply the previous results to a typical exemple of
an anomalous gauge theory: the chiral Schwinger model. The classical
action for this system
$$
   S_0(A_\mu;\psi,\bar\psi)=\int\dif^2 x \left[
   -\frac14 F^{\mu\nu}F_{\mu\nu}+i\bar\psi D\hspace{-2mm}\slash
   \frac{(1-\gamma_5)}{2}\psi
   \right],
$$
is invariant under the infinitesimal gauge transformations
$$
  \delta A_\mu=\partial_\mu\veps,\quad\quad
  \delta \psi= i\psi\veps,\quad\quad
  \delta  \bar\psi= -i\bar\psi\veps,
$$
where $\veps$ is the infinitesimal parameter of the abelian gauge
group.

After introducing a consistent regularization scheme
(e.g. Pauli-Villars regularization,\ldots), we obtain for the consistent
anomaly ${\cal A}$
\be
   {\cal A}(A_\mu)=\frac{i}{4\pi}
  \left[(1-a)\partial_\mu A^\mu-\veps^{\mu\nu}\partial_\mu A_\nu\right],
\label{an 2}
\ee
where $a$ is an arbitrary regularization parameter \cite{JR85}.

Let us now concentrate on the $M_1$ term.
As a path of integration in the general expression (\ref{wesszumino})
we can take
$$
   \theta'=\theta\cdot t,\quad\quad t\in[0,1],
$$
from which we obtain, taking into account that for the abelian gauge
group in consideration $\lambda^\alpha_\beta=1$,
\be
   M_1(A_\mu,\theta)=\frac1{4\pi}\int\dif^2 x\int_0^1\dif t
  \left[(1-a)\partial_\mu A^{'\mu}-\veps^{\mu\nu}\partial_\mu
   A'_\nu\right]\theta,
\label{m 1 c s}
\ee
where $A'_\mu=A'_\mu(t)$ is the gauge transformed of $A_\mu$ with
parameter $\theta\cdot t$, i.e.,
$A'_\mu(t)=(A_\mu+\partial_\mu\theta \cdot t)$. The susbstitution of
this expression in (\ref{m 1 c s}) and the integration
over the parameter $t$ yield (modulo total derivatives)
$$
   M_1(A_\mu,\theta)=\frac1{4\pi}\int\dif^2 x\left\{
  \frac{(a-1)}2\partial_\mu\theta\partial^\mu\theta-\theta
  \left[(a-1)\partial_\mu A^\mu+\veps^{\mu\nu}\partial_\mu
   A_\nu\right]\right\},
$$
which is exactly the Wess-Zumino action for this model.

It is easy to verify that the gauge variation of $M_1$ reproduces the
anomaly $\cal A$ \bref{an 2}
$$
  \delta M_1(A_\mu,\theta)=\int\dif^2 x\left[i \gh A\cdot \veps\right],
$$
with the gauge transformation for the extra field $\theta$ given by
$$
   \delta\theta=-\veps.
$$

Finally, the functional integration
$$
   \int\gh D \theta \exp\{i M_1(A_\mu,\theta)\}\sim
   \exp\{i \tilde M_1(A_\mu)\},
$$
yields the effective action
$$
  \tilde M_1(A_\mu)=\frac{1}{8\pi}\int\dif^2 x\left\{A_\mu
  \left[a g^{\mu\nu}-(g^{\mu\alpha}+\veps^{\mu\alpha})
  \left(\frac{\partial_\alpha\partial_\beta}{\Box}\right)
  (g^{\nu\beta}+\veps^{\nu\beta})\right]A_\nu\right\},
$$
which is non-local as corresponds to an anomalous gauge theory.

\section*{Acknowledgements}

\hspace{\parindent}%
The author would like to thank J.\,Gomis and J.\,M.\,Pons for
suggestions and critical reading of the manuscript. Useful discussions
with J.\,Roca are also acknowledged.

This work has been partially supported by a CICYT project
no.\,AEN89-0347.

\endsecteqno

\end{document}